\documentclass[prd,preprintnumbers,nofootinbib,twocolumn]{revtex4} 

\usepackage{graphicx}
\usepackage{dcolumn}
\usepackage{bm}
\usepackage{latexsym}
\usepackage{amsfonts}
\usepackage{amssymb}
\usepackage{amsmath}

\begin{document}

\preprint{IPMU09-0136}

\title{
Stellar center is dynamical in Ho\v{r}ava-Lifshitz gravity
}

\author{Keisuke Izumi}
\email{keisuke.izumi@ipmu.jp}
\author{Shinji Mukohyama}
\email{shinji.mukohyama@ipmu.jp}
\affiliation{
 IPMU, The University of Tokyo, Kashiwa, Chiba 277-8582, Japan}

\date{\today}

\begin{abstract}
 In Ho\v{r}ava-Lifshitz gravity, regularity of a solution requires
 smoothness of not only the spacetime geometry but also the
 foliation. As a result, the regularity condition at the center of a 
 star is more restrictive than in general relativity. 
 Assuming that the energy density is a piecewise-continuous,
 non-negative function of the pressure and that the pressure at the
 center is positive, we prove that the momentum conservation law is
 incompatible with the regularity at the center for any
 spherically-symmetric, static configurations. The proof is totally
 insensitive to the structure of higher spatial curvature terms and,
 thus, holds for any values of the dynamical critical exponent $z$. 
 Therefore, we conclude that a spherically-symmetric star should include
 a time-dependent region near the center. This supports the picture that
 Ho\v{r}ava-Lifshitz gravity does not recover general relativity at low
 energy but can instead mimic general relativity plus cold dark matter:
 the ``cold dark matter'' accretes toward a star and thus makes the
 stellar center dynamical. We also comment on the condition under which
 linear instability of the scalar graviton does not show up. 
\end{abstract}

\maketitle

\section{Introduction}
\label{sec:intro}

Recently, a power-counting renormalizable gravity theory was proposed by
Ho\v{r}ava~\cite{Horava:2009uw,Horava:2009if}. The essential reason for 
the power-counting renormalizability is that in the ultraviolet (UV), the
theory exhibits the Lifshitz-type anisotropic scaling 
\begin{equation}
 t \to b^z t,\  \vec{x} \to b \vec{x},
\end{equation}
with the dynamical critical exponent $z\geq 3$. Because of the Lifshitz
scaling, this theory is often called Ho\v{r}ava-Lifshitz
gravity. Although power-counting renormalizability does not necessarily
imply renormalizability, there is a good possibility that the theory is
unitary and renormalizable. Evidently, this is one of the driving forces
behind recent enthusiasms for research on various aspects of the theory
such as cosmology ~\cite{cos,DM}, black holes~\cite{BH}, the solar
system dynamics~\cite{sol}, and so on~\cite{others}.

In Ho\v{r}ava-Lifshitz gravity a black hole is not
``black'' for high-energy probes even classically while it is indeed
``black'' for low-energy probes. The Lifshitz scaling implies that the
dispersion relation is of the form $\omega^2\propto k^{2z}$ and the
group velocity $v_g\propto k^{z-1}$ is superluminal in the UV. This
means that,  although there are ``black hole'' solutions in this theory,
waves with sufficiently high frequencies can probe the region deep
inside the gravitational radius and the central singularity can be
seen. On the other hand, waves with low frequencies follow the ordinary
dispersion relation $\omega^2\propto k^2$ and thus cannot probe the
region inside the gravitational radius. In this sense a black hole
horizon is only an emergent concept in the infrared (IR).

In order to investigate what happens inside a black hole, 
especially near the singularity and to see if
the classical singularity can be resolved by quantum effects, we need to
formulate a regular initial condition and to evolve it dynamically
towards gravitational collapse.

The purpose of this paper is, as a first step towards this outstanding
problem, to initiate investigation of stellar solutions in
Ho\v{r}ava-Lifshitz gravity. This is important also for constraining the 
theory from astrophysical observations. A natural strategy would be to
try to construct a globally static solution regular at the
center. Surprisingly, we find a no-go result: we prove that a
spherically-symmetric, globally static solution can not be constructed
in Ho\v{r}ava-Lifshitz gravity under the assumption that the energy
density is a piecewise-continuous, non-negative function of the pressure
and that the pressure at the center is positive. Therefore, under the
assumption in the matter sector, a non-singular stellar solution should
include a time-dependent region, presumably near the center. This
supports the picture that Ho\v{r}ava-Lifshitz gravity does not recover
general relativity at low energy but can instead mimic general
relativity plus cold dark matter: the ``cold dark matter'' accretes
toward a star and thus makes the stellar center dynamical.

The rest of this paper is organized as follows. In section
\ref{sec:Horava} we shall describe the definition, properties and
problems of Ho\v{r}ava-Lifshitz theory. In section \ref{sec:proof} we
present the main result of this paper, i.e. the proof of non-existence
of spherically-symmetric, globally-static solution regular at the center
under a set of reasonable assumptions in the matter sector. Section
\ref{sec:summary} is devoted to a summary of this paper and discussion.
In appendix \ref{app:vac} we present asymptotically-flat vacuum
solutions with $\lambda=1$. Those solutions presented in appendix
\ref{app:vac} are used in appendix \ref{app:another-no-go}, where we
show yet another no-go result for globally-static star solutions with
$\lambda=1$. In appendix \ref{app:instability} we derive the condition
(\ref{eqn:cscond}) under which linear instability of the scalar graviton
does not show up. In appendix \ref{app:foliation} we consider a simple
stellar solution in general relativity in Painlev\'{e}-Gullstrand-like 
coordinate system to see how significant the regularity of time
foliation is and to foresee what happens near a stellar center in
Ho\v{r}ava-Lifshitz gravity.

\section{Properties of Ho\v{r}ava-Lifshitz gravity}
\label{sec:Horava}

In this section we shall describe the definition, properties and 
problems of Ho\v{r}ava-Lifshitz gravity.

\subsection{Definition and basic feature of Ho\v{r}ava-Lifshitz gravity}

Ho\v{r}ava-Lifshitz theory is fundamentally non-relativistic since the
Lifshitz scaling treats time and space differently. To be more precise,
the fundamental symmetry of the theory is invariance under the
foliation-preserving diffeomorphism: 
\begin{equation}
 t \to \tilde{t}(t), \quad x^i \to \tilde{x}^i(t,x). \label{eqn:symmetry}
\end{equation}
The foliation-preserving diffeomorphism consists of the space-independent
time reparametrization and the time-dependent spatial diffeomorphism. 
This means that the foliation of
the spacetime manifold by constant-time hypersurfaces is not a gauge but
has a physical meaning.

Basic quantities in the gravity sector are the lapse function $N(t)$,
the shift vector $N^i(t,\vec{x})$ and the three-dimensional spatial
metric $g_{ij}(t,\vec{x})$. These variables can be combined to form a 
four-dimensional metric in the Arnowitt-Deser-Misner (ADM) form:
\begin{equation}
 ds^2 = -N^2dt^2 + g_{ij}(dx^i+N^idt)(dx^j+N^jdt). 
  \label{eqn:ADMmetric}
\end{equation}
Since the lapse function is roughly speaking a gauge freedom associated
with the space-independent time reparametrization, it is rather natural
to restrict the lapse function to be space-independent. This condition,
called the projectability condition, is not only natural but also 
mandatory, as pointed out in Ho\v{r}ava's original 
paper~\cite{Horava:2009uw}. Indeed, if we abandoned the projectability
condition then we would face phenomenological
obstacles~\cite{Charmousis:2009tc} and theoretical
inconsistencies~\cite{Li:2009bg}. On the other hand, with the
projectability condition (and without the detailed balance condition),
the theory is free from those problems~\cite{DM}. (See the discussion about 
strong coupling in the next subsection.) Therefore,
throughout this paper, we impose the projectability condition and demand
that the lapse function be space-independent. The Hamiltonian constraint
is, as a result, not a local equation satisfied at each spatial point
but an equation integrated over a whole space.

Under the infinitesimal transformation
%
\begin{equation}
 \delta t = f(t), \quad \delta x^i = \zeta^i(t,x),
\end{equation}
$g_{ij}$, $N^i$ and $N$ transform as
%
\begin{eqnarray}
 \delta g_{ij} & = & f\partial_t g_{ij} + {\cal L}_{\zeta}g_{ij}
  \nonumber\\
 \delta N^i & = & \partial_t (N^i f) + \partial_t \zeta^i
  + {\cal L}_{\zeta}N^i, \nonumber\\
 \delta (N_i) & = & \partial_t (N_i f) + g_{ij}\partial_t \zeta^j
  + {\cal L}_{\zeta}N_i, \nonumber\\
 \delta N & = & \partial_t(N f). 
  \label{eqn:infinitesimal-tr}
\end{eqnarray} 
Thus, $N$ remains independent of spatial coordinates after the
transformation. In the IR, where $dt$ and $dx^i$ have the same scaling
dimension, it makes sense to assemble $g_{ij}$, $N^i$ and $N$ into a
$4$-dimensional metric in the ADM form (\ref{eqn:ADMmetric}).  The
action is 
\begin{eqnarray}
 I & = &I_g+I_m, \\
 I_g & = & \frac{M_{Pl}^2}{2}
  \int dt dx^3N\sqrt{g}(K^{ij}K_{ij}-\lambda K^2\nonumber\\
 & & \qquad\qquad\qquad\qquad\qquad +\Lambda+R+L_{z>1}),
  \label{eqn:gravaction}
\end{eqnarray}
where
\begin{equation}
 K_{ij} = \frac{1}{2N}(\partial_t g_{ij}-D_iN_j-D_jN_i), \quad
 K = g^{ij}K_{ij}, 
\end{equation}
$D_i$ is the covariant derivative compatible with $g_{ij}$, 
$\Lambda$ is a cosmological constant, $R$ is the
Ricci scalar of $g_{ij}$, $L_{z>1}$ represents higher spatial curvature
terms and $I_m$ is the matter action. Here, we have rescaled the time
coordinate so that the coefficients of $K^{ij}K_{ij}$ and $R$ agree. 
Note that not only the gravitational action $I_g$ but also
the matter action $I_m$ should be invariant under the
foliation-preserving diffeomorphism.

By variation of the action with respect to $N(t)$, we obtain the
Hamiltonian constraint 
%
\begin{equation}
 H_{g\perp}+H_{m\perp}=0, \label{eqn:HamiltonianConstraint}
\end{equation}
where
%
\begin{eqnarray}
 H_{g\perp} & \equiv & -\frac{\delta I_g}{\delta N}
  = \int dx^3\sqrt{g} {\cal H}_{g\perp}, 
  \nonumber\\
 H_{m\perp} & \equiv & -\frac{\delta I_m}{\delta N}
  = \int dx^3\sqrt{g}\ T^{\perp}_{\perp}, 
\end{eqnarray}
and
%
\begin{eqnarray}
  {\cal H}_{g\perp} & = & \frac{M_{Pl}^2}{2}
  (K^{ij}p_{ij}-\Lambda-R-L_{z>1}),  \nonumber\\
 T^{\perp}_{\perp} & = & T_{\mu\nu}n^{\mu}n^{\nu}.
\end{eqnarray}
Here, $p_{ij}$ and $n^{\mu}$ are defined as 
%
\begin{equation}
 p_{ij} \equiv K_{ij} - \lambda Kg_{ij},
\end{equation}
and
%
\begin{equation}
 n_{\mu}dx^{\mu} = -Ndt, \quad
  n^{\mu}\partial_{\mu}= \frac{1}{N}(\partial_t-N^i\partial_i). 
  \label{eqn:unitnormal}
\end{equation}
Variation with respect to $N^i(t,x)$ leads to the momentum constraint 
%
\begin{equation}
 {\cal H}_{g i}+{\cal H}_{m i}=0, \label{eqn:MomentumConstraint}
\end{equation}
where
%
\begin{eqnarray}
 {\cal H}_{g i} & \equiv & -\frac{1}{\sqrt{g}}\frac{\delta I_g}{\delta N^i}
  = -M_{Pl}^2D^jp_{ij}, \nonumber\\
 {\cal H}_{m i} & \equiv & -\frac{1}{\sqrt{g}}\frac{\delta I_m}{\delta N^i}
  = T_{i\mu}n^{\mu}. 
\end{eqnarray}
Note that the momentum constraint is determined solely by the kinetic
terms and thus is totally insensitive to the structure of higher spatial
curvature terms. In particular, for $\lambda=1$ the momentum constraint
agrees with that in general relativity.

As in general relativity, the gravitational action can be written as the
sum of kinetic terms and constraints up to boundary terms:  
%
\begin{eqnarray}
 I_g & = & \int dt dx^3
  \left[\pi^{ij}\partial_tg_{ij}-N^i{\cal H}_{gi}\right]
  \nonumber\\
 & & - \int dt NH_{g\perp} + (\mbox{boundary terms}),
\end{eqnarray}
where $\pi^{ij}$ is momentum conjugate to $g_{ij}$ given by 
%
\begin{equation}
 \pi^{ij} \equiv \frac{\delta I_g}{\delta (\partial_tg_{ij})}
  = M_{Pl}^2\sqrt{g}p^{ij}, \quad
  p^{ij} \equiv g^{ik}g^{jl}p_{kl}. 
\end{equation}
The Hamiltonian corresponding to the time $t$ is the sum of constraints
and boundary terms as
%
\begin{equation}
 H_g[\partial_t] = NH_{g\perp} + \int dx^3 N^i{\cal H}_{gi}
  + (\mbox{boundary terms}). 
\end{equation}

Finally, by variation with respect to $g_{ij}(t,x)$, we obtain dynamical
equation 
%
\begin{equation}
 {\cal E}_{g ij}+{\cal E}_{m ij}=0, \label{eqn:DynamicalEquation}
\end{equation}
%
\begin{eqnarray}
 {\cal E}_{g ij} & \equiv & g_{ik}g_{jl}\frac{2}{N\sqrt{g}}
  \frac{\delta I_g}{\delta g_{kl}}, \nonumber\\
 {\cal E}_{m ij} & \equiv & g_{ik}g_{jl}\frac{2}{N\sqrt{g}}
  \frac{\delta I_m}{\delta g_{kl}}
  = T_{ij}. 
\end{eqnarray} 
The explicit expression for ${\cal E}_{g ij}$ is given by 
%
\begin{eqnarray}
 {\cal E}_{g ij} & = & M_{Pl}^2
  \left[
  -\frac{1}{N}(\partial_t-N^kD_k)p_{ij}
  \right.\nonumber\\
 & & + \frac{1}{N}(p_{ik}D_jN^k+p_{jk}D_iN^k)
  - Kp_{ij} + 2K_i^kp_{kj}
  \nonumber\\
 & & \left.
      + \frac{1}{2}g_{ij}K^{kl}p_{kl}+ \frac{1}{2}\Lambda g_{ij} 
      - G_{ij}\right]
 + {\cal E}_{z>1 ij},
\end{eqnarray} 
where ${\cal E}_{z>1 ij}$ is the contribution from $L_{z>1}$ and
$G_{ij}$ is Einstein tensor of $g_{ij}$.

The invariance of $I_{\alpha}$ under the infinitesimal transformation 
(\ref{eqn:infinitesimal-tr}) leads to the following conservation
equations, where $\alpha$ represents $g$ or $m$. 
%
\begin{eqnarray}
 0 & = & N\partial_t H_{\alpha\perp} \nonumber\\ 
 & & + \int dx^3\left[ N^i\partial_t(\sqrt{g}{\cal H}_{\alpha i})
  +\frac{1}{2}N\sqrt{g}{\cal E}_{\alpha}^{ij}\partial_tg_{ij}\right], 
 \\
 0 & = & \frac{1}{N}(\partial_t-N^jD_j){\cal H}_{\alpha i}
  + K{\cal H}_{\alpha i} \nonumber\\
 & & - \frac{1}{N}{\cal H}_{\alpha j}D_iN^j
  - D^j{\cal E}_{\alpha ij}. 
  \label{eqn:conservation}
\end{eqnarray} 

\subsection{Properties and problems of Ho\v{r}ava gravity}
\label{subsec:property}

\emph{Renormalization group (RG) flow of $\lambda$}\\
The IR limit of the theory is characterized by the parameter
$\lambda$. When $\lambda=1$, the gravitational action in
the IR limit is identical to the ADM form of the Einstein-Hilbert action  
except that the lapse function is independent of spatial coordinates. 
Note, however, that setting $\lambda=1$ at all scales would lead to a
problem since deviation from $\lambda=1$ in the UV is essential for
avoidance of codimension-one caustics~\cite{DM}. While the
(RG) flow of Ho\v{r}ava-Lifshitz gravity has not
yet been analyzed, we therefore suppose that $\lambda=1$ is an IR fixed
point of the RG flow so that $\lambda$ is sufficiently close to $1$ in
the IR but deviates from $1$ in the UV.  
\\

\emph{Instabilities}\\
In order to avoid ghost instability of the scalar graviton, the
coefficient of the time kinetic term $(\lambda-1)/(3\lambda-1)$ must be 
positive~\cite{Horava:2009uw}. This makes the sound speed squared
$c_s^2=-(\lambda-1)/(3\lambda-1)$
negative~\cite{Wang:2009yz} and leads to an IR
instability in the linear level. As shown in
Appendix~\ref{app:instability}, this type of instability does not show
up if 
\begin{equation}
 |c_s| < \max \left[ HL, (ML)^{-1}, |\Phi|^{1/2}\right],
  \label{eqn:cscond}
\end{equation}
where $M$ is the energy scale suppressing higher-derivative terms, $L$
($>0.01mm$) is the length scale of interest, $H$ is the Hubble expansion
rate at the time of interest and $\Phi$ is the Newtonian potential at
the position of interest. This is a condition on the way how $\lambda$
depends on $L$, $H$ and $\Phi$ and, thus, validity of this condition can
in principle be checked by analyzing the RG flow.  
\\

\emph{Strong coupling}\\
The strong coupling between the scalar graviton and matter pointed out
in \cite{Charmousis:2009tc} in the limit $\lambda\to 1$ is absent if the
projectability condition is imposed~\cite{DM}. On the other hand, even
with the projectability condition, there still remains the strong
self-coupling of the scalar graviton in the limit $\lambda\to 1$, as is
clear from Ho\v{r}ava's original paper~\cite{Horava:2009uw}. However,
this is not a problem if the scalar graviton decouples from the rest of
the world at the nonlinear level. In massive gravity
theories~\cite{Fierz:1939ix} this kind of decoupling phenomenon due to
nonlinear dynamics is known as Vainstein
effect~\cite{Vainshtein:1972sx}. A potential problem is that the strong
self-coupling makes quantum corrections to the classical action very
large. In non-renormalizable theories like massive gravity, this can be
a fatal flaw since we really need to deal with an unknown quantum action
including quantum gravity effects while the Vainstein effect has been
shown only for a classical action with finite number of terms. On the
other hand, in Ho\v{r}ava-Lifshitz gravity, if the theory is
renormalizable then we can safely use the renormalizable action with
finite number of terms to investigate whether the scalar graviton really
decouples or not. Detailed investigation of decoupling of the scalar
graviton is one of the most important issues, but the present paper aims
to shed light on a different aspect of the theory. 
\\

\emph{Black hole}\\
With $\lambda=1$ and vanishing cosmological constant, Schwarzschild
spacetime is locally an exact vacuum solution of Ho\v{r}ava-Lifshitz
gravity (See appendix \ref{app:vac}). In the Painlev\'{e}-Gullstrand coordinate system the
Schwarzschild metric is 
\begin{equation}
 ds^2 = -dt^2 + \left(dr\pm \sqrt{\frac{2M}{r}}dt\right)^2
  + r^2d\Omega_2^2, \label{eqn:Schwarzschild}
\end{equation}
and satisfies the projectability condition. Here, $d\Omega_2^2$ is the
metric of the unit sphere. Since the spatial metric $dr^2+r^2d\Omega^2_2$
is flat, higher spatial curvature terms do not change the solution at all. 

As we note in Section~\ref{sec:intro}, 
in this theory a black hole is not ``black'' in the UV 
and we can see the central singularity of a Schwarzschild black hole 
if it exists. 
Since at the singular point the basic equations~(\ref{eqn:MomentumConstraint}) 
and (\ref{eqn:DynamicalEquation}) are not satisfied, 
a Schwarzschild metric must be modified near the center.
For the Schwarzschild metric (\ref{eqn:Schwarzschild}), the extrinsic
curvature of constant-time hypersurfaces becomes large near the center
and the system enters the UV regime. Therefore, $\lambda$ should deviate
from $1$ and the Schwarzschild spacetime is no more a valid 
description of the geometry near the center. For this reason, the
apparent singularity at the center is not physical.

\section{No spherically symmetric and static solution} 
\label{sec:proof}

As the last topic of Section~\ref{sec:Horava}, we touched upon the
problem of the central singularity in a black hole. In order to
investigate what happens inside the black hole, we need to study
gravitational collapse and formation of a black hole. For this purpose
we need to formulate a regular initial condition. As a first step, in 
the present paper we consider a static star. Surprisingly enough, we
find a no-go result: a spherically-symmetric, globally static solution
can not be constructed in Ho\v{r}ava-Lifshitz gravity under the
assumption that the energy density is a piecewise-continuous,
non-negative function of the pressure and that the pressure at the
center is positive. This section provides a proof of this statement.

\subsection{Painlev\'{e}-Gullstrand coordinate}

We consider spherical-symmetric and static configurations. Since the
lapse function does not depend on spatial coordinates, we can set it to
unity by space-independent time reparametrization: 
\begin{equation}
 N = 1.
\end{equation}
For a spherically symmetric, static configuration, we can express the
shift vector and the spatial metric as 
\begin{equation}
 N^i\partial_i = \beta(x)\partial_x, \quad
 g_{ij}dx^idx^j = dx^2 + r^2(x)d\Omega^2_2,
\end{equation}
where $d \Omega_2^2$ is the metric of the unit sphere and $x$ is the
proper distance from the center. Non-vanishing components of the Ricci
tensor and Ricci scalar for the three-dimensional geometry are
\begin{eqnarray}
 R^x_x & = & -\frac{2r''}{r}, \quad
  R^{\theta}_{\theta} =
  \frac{1}{r^2}\left[1-rr''-(r')^2\right],\\
 R & = & \frac{2}{r^2}\left[1-2rr''-(r')^2\right],
 \label{eqn:R}
\end{eqnarray}
and the extrinsic curvature and its trace are
\begin{equation}
 K_{ij}dx^idx^j = -\beta'dx^2 
  -\beta rr' d\Omega^2_2, \quad
  K = -\frac{(r^2\beta)'}{r^2},
  \label{eqn:extrinsic-curvature}
\end{equation}
where a prime denotes derivative w.r.t. $x$. The corresponding ADM
metric is 
\begin{equation}
 g^{(4)}_{\mu\nu}dx^{\mu}dx^{\nu}
  = -dt^2  + [dx+\beta(x)dt]^2 + r^2(x) d\Omega^2_2. 
\end{equation}
This is an analogue of the Painlev\'{e}-Gullstrand coordinate
system. Note that 
\begin{equation}
 \xi^{\mu} = \left(\frac{\partial}{\partial t}\right)^{\mu}
  \label{eqn:Killing}
\end{equation}
is a timelike Killing vector. Global staticity requires $\xi^{\mu}$ to
be globally timelike, i.e. $1-\beta^2>0$ everywhere. The unit vector
$n^{\mu}$ normal to the constant time hypersurface is given by 
\begin{eqnarray}
 n_{\mu}dx^{\mu} = -dt, \quad
  n^{\mu}\partial_{\mu} = \partial_t - \beta\partial_x. 
\end{eqnarray}

It is instructive to see that the Newton potential $\phi(r)$ and the
Misner-Sharp energy $m(r)$ are written as 
\begin{eqnarray}
 e^{2\phi(r)}  & = & -g^{(4)}_{\mu\nu}\xi^{\mu}\xi^{\mu}
  = 1-\beta^2, \\
 1-\frac{2m(r)}{r}  & = & g^{(4)\mu\nu}(dr)_{\mu}(dr)_{\nu}
  = (1-\beta^2)(r')^2. 
\end{eqnarray}

\subsection{Matter sector}

As for (real) matter, for simplicity we consider the perfect-fluid form
which is at rest w.r.t. the Killing vector $\xi^\mu$:
\begin{eqnarray}
 T_{\mu\nu} & = & \rho(x) u_{\mu}u_{\nu} +
  P(x)\left[ g^{(4)}_{\mu\nu} + u_{\mu}u_{\nu} \right], 
  \label{eqn:Tmunu} \\
 u^{\mu} & = & \frac{\xi^{\mu}}{\sqrt{1-\beta^2}}. 
\end{eqnarray}
Its components relevant for the ADM decomposition are 
\begin{eqnarray}
 T_{\mu\nu}n^\mu n^\nu & = & (1-\beta^2)^{-1}(\rho + P) - P,\\
 T_{\mu i}n^\mu dx^i & = & (1-\beta^2)^{-1}(\rho + P)\beta dx,\\
 T_{ij}dx^idx^j & = & [\beta^2 \rho + P] dx^2 + Pr^2d\Omega^2_2.
\end{eqnarray}

\subsection{Inconsistency near the center}
\label{subsec:inconsistency}

In this subsection, we show that a spherically-symmetric compact object
cannot be globally-static under a set of reasonable assumptions in the
matter sector. We assume that the energy density $\rho$ is a
piecewise-continuous~\footnote{
We assume piecewise-continuity instead of continuity, in order to allow
$dP/d\rho$ to vanish in a finite interval of $\rho$. 
} 
non-negative function of the pressure $P$ and that the pressure at the
center $P_c$ is positive. The three-dimensional spatial geometry and the
extrinsic curvature must be regular at the center because the
constant-time surfaces are physically embeded in Ho\v{r}ava-Lifshitz
gravity. (See Appendix \ref{app:foliation} on this point.) We prove that the
regularity condition at the center is incompatible with the momentum
conservation law of the matter under the above assumptions. Since the
momentum conservation equation does not include higher spatial curvature
terms in the gravity action, our proof is totally insensitive to the
structure of higher spatial curvature terms and holds for any values of
the dynamical critical exponent $z$.

In the following argument, we make the proposition that there exists a 
spherically-symmetric, globally-static, regular solution, and show
contradiction. As commented after eq.~(\ref{eqn:Killing}), the
global-staticity implies that $(1-\beta^2)$ is positive everywhere.

The momentum conservation equation, (\ref{eqn:conservation}) with
$\alpha=m$, becomes 
\begin{eqnarray}
 P'(1-\beta^2)+(\rho+P)(1-\beta^2)'=0.
\label{momconserve2}
\end{eqnarray}
The regularity of the extrinsic curvature
(\ref{eqn:extrinsic-curvature}) implies that $\beta'$ is finite. 
This and (\ref{momconserve2}) imply that $P'$ is also finite. As a
corollary, $\beta$ and $P$ are continuous functions of $x$. Since $\rho$
is assumed to be a piecewise continuous function of $P$, this means that
$\rho+P$ is a piecewise continuous function of $x$.

Let $x_c$ be the value of $x$ at the center. Since we have assumed that
$\rho$ is non-negative everywhere and that $P_c>0$, the continuity of
$P(x)$ implies that $\rho+P$ is positive in a neighborhood of the
center. Now let us define $x_0$ as the minimal value for which at least
one of  $(\rho+P)|_{x=x_0}$, $\lim_{x\to x_0-0}(\rho+P)$ and 
$\lim_{x\to x_0+0}(\rho+P)$ is non-positive.

Dividing eq.~(\ref{momconserve2}) by $(\rho+P)(1-\beta^2)$ and
integrating it over the interval $x_c\leq x<x_0$, we obtain
\begin{eqnarray}
\ln(1-\beta_0^2)-\ln(1-\beta_c^2)
=-\int^{x_0-0}_{x_c} \frac{P'}{\rho+P}dx,
\label{momconserve}
\end{eqnarray}
where $\beta_c\equiv\beta(x=x_c)$ and $\beta_0\equiv\beta(x=x_0)$. 
The regularity of the Ricci scalar (\ref{eqn:R}) and the extrinsic
curvature (\ref{eqn:extrinsic-curvature}) at the center implies that
${r'}_c=1$ and $\beta_c=0$, where ${r'}_c$ is the value of $r'$ at the
center. Therefore, the left hand side of eq.~(\ref{momconserve}) is
non-positive.

Since $P$ is a differentiable function of $x$, the right hand side of 
eq.~(\ref{momconserve}) can be transformed as 
\begin{eqnarray}
-\int^{x_0-0}_{x_c} \frac{P'}{\rho+P}dx
=-\int^{P_0}_{P_c}\frac{dP}{\rho(P)+P},
\label{Pint}
\end{eqnarray}
where $P_0\equiv P(x=x_0)$. The definition of $x_0$ implies that at
least one of $(\rho+P)|_{x=x_0}$, $\lim_{x\to x_0-0}(\rho+P)$ and 
$\lim_{x\to x_0+0}(\rho+P)$ is non-positive. Since we have assumed that
$\rho$ is non-negative everywhere, 
$P_0=\lim_{x\to x_0-0}P=\lim_{x\to x_0+0}P$ is non-positive. Thus, we
have 
\begin{eqnarray}
 P_0 \leq 0 < P_c. 
\end{eqnarray}
This implies positivity of the right hand side of (\ref{Pint}) 
since from the definition of $x_0$ the integrand is positive 
in the domain of integration.
This leads to an contradiction 
with the previous statement that the left hand side of 
(\ref{momconserve}) should be non-positive.

\section{Summary and Discussion}
\label{sec:summary}

In Ho\v{r}ava-Lifshitz gravity, regularity of a solution requires
smoothness of not only the spacetime geometry but also the foliation. As
a result, the regularity condition at the center of a star is more
restrictive than in general relativity. Under the assumptions 
that the energy density is piecewise-continuous non-negative function of
the pressure and that the pressure at the center is positive, we have
proved that the momentum conservation law is incompatible with the
regularity at the center for any spherically-symmetric, globally-static
configurations. The proof is totally insensitive to the structure of
higher spatial curvature terms and, thus, holds for any values of the
dynamical critical exponent $z$. Therefore, under the assumption we made
on the matter sector, we conclude that a spherically-symmetric star
should include a time-dependent region, presumably near the center. 

The assumptions we made are physically natural. For example, a
polytropic fluid satisfies them. Note that our proof does not assume
asymptotic flatness and that a cosmological constant $\Lambda$ can be
included in the gravity action (\ref{eqn:gravaction}). Shifting $\rho$,
$-P$ and $-M_{Pl}^2\Lambda$ with the same amount does not change the
physical system but may validate/invalidate some of the assumptions of
the proof. In order to construct a static star solution, we need to
violate at least one of the assumptions for all possible choices of such
a shift. One possibility is to introduce an exotic matter such as a
quintessence field. Introduction of an exotic matter is, however, not
necessarily sufficient for the existence of a static star solution.

One must not consider our result, i.e. nonexistence of static star, as a
serious problem of Ho\v{r}ava-Lifshitz theory. It is known that this
theory does not recover general relativity at low energy but can instead
mimic general relativity plus cold dark matter~\cite{DM}~\footnote{The
constraint algebra is smaller than in general relativity since the time
slicing is synchronized with the rest frame of cold dark matter in the
theory level.} The existence of built-in ``cold dark matter'' is an
inevitable prediction of the theory and might solve the mystery of dark
matter in the universe. Our result in the present paper is totally
consistent with this picture: as in the standard cold dark matter
scenario, the ``cold dark matter'' accretes toward a star and thus
inevitably makes the stellar center dynamical. This is the physical
reason why there is no static star in Ho\v{r}ava-Lifshitz theory and,
thus, our result strongly supports the ``dark matter as an integration
constant'' scenario~\cite{DM}.

In subsection \ref{subsec:property}, we have commented on the Vainstein
effect for the scalar graviton, and also derived the condition
(\ref{eqn:cscond}) under which linear instability of the scalar graviton
does not show up. This condition should be considered as a
phenomenological constraint on properties of the renormalization group
(RG) flow. Detailed analysis of the RG flow and the fate of scalar
graviton will be one of important future subjects.

\section*{Acknowledgments}

The authors would like to thank K.~Takahashi for useful discussions. 
The work of S.M. was supported in part by 
Grant-in-Aid for Young Scientists (B) No.~17740134, 
Grant-in-Aid for Creative Scientific Research No.~19GS0219,
Grant-in-Aid for Scientific Research on Innovative Areas No.~21111006, 
Grant-in-Aid for Scientific Research (C) No.~21540278, and 
the Mitsubishi Foundation. 
This work was supported by World Premier International Research Center
Initiative (WPI Initiative), MEXT, Japan. 

\appendix

\section{Asymptotically-flat vacuum solution with $\lambda=1$ 
and without cosmological constant}
\label{app:vac}

For $T_{\mu\nu}=0$, $\Lambda=0$ and $\lambda=1$, the momentum constraint
(\ref{eqn:MomentumConstraint}) and the dynamical equation
(\ref{eqn:DynamicalEquation}) become 
\begin{eqnarray}
 0 & = & \beta\frac{r''}{r},\label{eqn:vacmom} \\
 0 & = & \frac{1}{r^2}
  \left[2\beta r r''-(1-\beta^2)(r')^2 \right.
  \nonumber\\
 & & \left. \quad + 2 \beta\beta'r r' +1\right]
  + \frac{1}{2}(r')^2{\cal E}_{z>1}^{xx}, \label{eqn:vacrr}
\end{eqnarray}
where ${\cal E}_{z>1}^{ij}\equiv g^{ik}g^{jl}{\cal E}_{z>1 kl}$.

On the other hand, we shall not impose the Hamiltonian constraint
equation (\ref{eqn:HamiltonianConstraint}) to our solutions. The reason
is as follows. In Ho\v{r}ava-Lifshitz gravity the Hamiltonian constraint
is not a local equation satisfied at each spatial point but an equation
integrated over a whole space. On the other hand, our assumption that
$T_{\mu\nu}=0$ is justified only if we consider physics at length scales
sufficiently shorter than the cosmological horizon. Otherwise, the
stress-energy tensor of cosmological fluids should be included. Also,
our assumption of staticity completely neglects cosmological expansion
of our universe. Therefore, static vacuum solutions would be valid only
in a (large but) finite region. For this reason, it is not appropriate
to substitute a static vacuum solution to the global Hamiltonian
constraint, which is an integral over a whole space including the region
outside the cosmological horizon. The contribution from the region
within the regime of validity of a static vacuum solution to the
integral can easily be canceled by the contribution from the region
outside the regime of validity of the solution. Therefore, we must not
impose the Hamiltonian constraint to our static vacuum solution.

There are two branches of solutions to the momentum constraint equation
(\ref{eqn:vacmom}): $\beta=0$ and $r'=r_1$ (constant). In the following
we shall consider each branch separately. Without loss of generality, we
assume that $r_1\geq 0$. (If $r_1<0$ then we can make redefinition 
$x\to -x$ so that $r_1\to -r_1$.)

Let us consider the first branch, $\beta=0$. In this case the dynamical
equation (\ref{eqn:vacrr}) is reduced to 
\begin{equation}
 0 = \frac{1}{r^2}[1-(r')^2]  + \frac{1}{2}(r')^2{\cal E}_{z>1}^{xx}.
 \label{beta=0}
\end{equation}
For asymptotically-flat solutions, the first term behaves as 
$\sim 1/r^2$ unless $r'=\pm 1$, while ${\cal E}_{z}$ decays at least as
fast as $1/r^4$ in the asymptotic region. Therefore, asymptotically flat
solutions with $\beta=0$ has $r'=\pm 1$ and can be included in the
second branch of solutions to (\ref{eqn:vacmom}), i.e. $r'=r_1$
(constant).

In the second branch, $r'=r_1$ (constant), the Ricci tensor and the
Ricci scalar become 
\begin{equation}
 R^x_x=0,\quad R^\theta_\theta={1-r_1^2\over r^2}, \quad
  R={2-2r_1^2\over r^2}. \label{eqn:r1R}
\end{equation}
Evidently, the value $r_1=1$ is special. In this case, all components
of the Ricci tensor vanish and, thus, ${\cal E}_{z>1}^{xx}=0$. Thus, for 
$r_1=1$, (\ref{eqn:vacrr}) is reduced to $(r\beta^2)'=0$ and leads to
the Schwarzschild solution (\ref{eqn:Schwarzschild}).

For $r'=r_1$ (constant) with $r_1\ne 1$, we can see from
eq.~(\ref{eqn:r1R}) that those terms in ${\cal E}_{z>1}^{xx}$ including
$2n$ spatial derivatives are proportional to $1/r^{2n}$, where
$n=2,\cdots,z$. Thus, the dynamical equation (\ref{eqn:vacrr}) is
written as 
\begin{equation}
 r_1(\beta^2 r)'=r_1^2-1-\sum_{n=2}^z {\alpha_n(r_1) \over r^{2n}}, 
  \label{eqn:betaeq}
\end{equation}
where $\alpha_n(r_1)$ ($n=2,\cdots,z$) are constants depending on the
constant $r_1$ and the structure of the higher spatial curvature terms
$I_{z>1}$. Since all components of the Ricci tensor vanish for $r_1=1$,
we have
\begin{equation}
 \alpha_n(1) = 0, \quad (n=2,\cdots,z). \label{eqn:alpha1vanish}
\end{equation}
Integrating the equation (\ref{eqn:betaeq}), we obtain
\begin{eqnarray}
 \beta^2 & = & \frac{r_1^2-1}{r_1^2} +\frac{2\mu}{r}
  + \sum_{n=2}^z \frac{\alpha_n(r_1)}{2n-1} \frac{1}{r^{2n}},
  \label{eqn:vacsolbeta}
\end{eqnarray}
where $\mu$ is an integration constant.

Note that $\beta$ must be real and thus $\beta^2\geq 0$. Since
(\ref{eqn:vacsolbeta}) implies that $\beta^2\to (r_1^2-1)/r_1^2$
($r\to\infty$), it follows that $r_1\geq 1$. (Note that we have assumed
that $r_1\geq 0$ without loss of generality.)

In summary, the asymptotically-flat vacuum solution is characterized by
two integration constants $r_1$ ($\geq 1$) and $\mu$. The solution is
given by $r'=r_1$ and (\ref{eqn:vacsolbeta}). When $r_1=1$, the solution
is reduced to the Schwarzschild spacetime (\ref{eqn:Schwarzschild}) with
$M=\mu$ since $\alpha_n(1)$ vanishes as shown in
(\ref{eqn:alpha1vanish}).

\section{Yet another no-go result}
\label{app:another-no-go}

In this appendix we show yet another no-go result, using the momentum
constraint equation. As in section \ref{sec:proof}, we make the
proposition that there exists a spherically-symmetric, globally-static,
regular solution under a set of assumptions, and show contradiction. We
set $\lambda=1$ and restrict our consideration to asymptotically flat
solutions (thus with vanishing cosmological constant). We also assume
that the shift $\beta$ is analytic. As for the matter sector, we only
assume the null energy condition.

The momentum constraint equation is 
\begin{equation}
 \frac{\beta}{r}
  \left[r''+\frac{r}{2M_{Pl}^2}\frac{\rho+P}{1-\beta^2}\right]
  = 0
\end{equation}
and is insensitive to the structure of higher spatial curvature terms. 
If $\beta=0$ in a finite interval of $x$ then $\beta=0$ everywhere
since $\beta$ is assumed to be analytic. This corresponds to a trivial
flat solution. We therefore suppose that 
\begin{equation}
 r''+\frac{r}{2M_{Pl}^2}\frac{\rho+P}{1-\beta^2} = 0
 \label{eqn:mom}
\end{equation}
is satisfied everywhere. If the null energy condition ($\rho+P\geq 0$)
is satisfied then by integrating (\ref{eqn:mom}) from the center towards
outside we obtain 
\begin{equation}
 r'|_{x=x_{out}}-r'|_{x=x_c} = -\int_{x_c}^{x_{out}}
  \frac{r}{2M_{Pl}^2}\frac{\rho+P}{1-\beta^2}dx \leq 0,
\end{equation}
where $x=x_c$ is the center and $x_{out}>x_c$. The equality holds if and
only if $\rho+P=0$ everywhere in the region $x_c\leq x\leq x_{out}$. 
Note that, as already stated just after (\ref{eqn:Killing}), the global
staticity requires $1-\beta^2>0$. Since we are interested in a stellar
solution, we suppose that $\rho+P\neq 0$ somewhere. Therefore, we obtain 
\begin{equation}
 r'|_{x=x_{out}} < r'|_{x=x_c}, \quad \mbox{for sufficiently large }
  x_{out}. 
  \label{eqn:drdx-decrease}
\end{equation}
Now we can show that the regularity of the solution at the center is
incompatible with the asymptotic flatness. The regularity at the center
requires that $r'|_{x=x_c}=1$. On the other hand, in Appendix~\ref{app:vac} 
we see that the asymptotically-flat vacuum
solution always has $r'\geq 1$. Therefore, the stellar solution can be 
connected to the vacuum solution in the asymptotic region only if 
$r'|_{x=x_{out}}\geq 1$, where we have chosen a sufficiently large
$x_{out}$. This is in conflict with (\ref{eqn:drdx-decrease}).

\section{Stealth linear instability of scalar graviton}
\label{app:instability}
 
As shown in \cite{DM}, the lack of local Hamiltonian constraint leads to
``dark matter as an integration constant'', a non-dynamical component
which behaves like pressure-less dust. As in the standard CDM scenario, 
the dust-like component exhibits Jeans instability and forms large-scale
structures in the universe. The timescale of Jeans instability is 
\begin{equation}
 t_J \sim (G_N\rho)^{-1/2}, 
\end{equation}
where $G_N$ is Newton constant and $\rho$ is the energy density at the
position of interest. Note that this instability is necessary for
structure formation if we consider the dust-like component as an
alternative to CDM.

As mentioned in the introduction, the scalar graviton exhibits linear
instability due to the negative sound speed squared,
$c_s^2=-(\lambda-1)/(3\lambda-1)<0$. The corresponding time scale is
\begin{equation}
 t_L \sim \frac{L}{|c_s|},
\end{equation}
where $L$ is the length scale of interest. Thus, as far as 
\begin{equation}
 t_L > t_J, \label{eqn:tLtJ}
\end{equation}
the linear instability does not show up.

Since the dispersion relation is of the form 
\begin{equation}
 \omega^2 = k^2\times \left[c_s^2 + O(k^2/M^2)\right],
  \label{eqn:higher-derivative}
\end{equation}
the linear instability is stabilized by higher derivative terms if
\begin{equation}
 |c_s| < \frac{1}{ML}, \label{eqn:hdterm}
\end{equation}
where $M$ is the energy scale suppressing higher derivative terms and we
have assumed that the coefficient of $k^2/M^2$ in the square bracket in
(\ref{eqn:higher-derivative}) is positive and of order unity. Also, the
linear instability is tamed by Hubble friction if 
\begin{equation}
 t_L > H^{-1}, \label{eqn:HtL}
\end{equation}
where $H$ is the Hubble expansion rate at the time of interest.

In summary, if one or more of the three conditions (\ref{eqn:tLtJ}),
(\ref{eqn:hdterm}) and (\ref{eqn:HtL}) is satisfied then the linear
instability of the scalar graviton does not show up. By introducing
Newton potential $\Phi\sim -G_N\rho L^2$, these conditions are
summarized as 
\begin{equation}
 |c_s| < \max \left[ HL, (ML)^{-1}, |\Phi|^{1/2}\right].  \label{eqn:csbound}
\end{equation}
As argued in \cite{DM}, nonlinear extension of the linear instability
should be formation of would-be caustics and bounce. For length scales
shorter than $\sim 0.01mm$, we do not experimentally know how gravity
behaves and, thus, the existence of formation of would-be caustics and
bounce does not contradict with any experiments. In some sense, this is
similar to the so called spacetime foam. Therefore, in
(\ref{eqn:csbound}), we do not have to consider the length scale $L$
shorter than $\sim 0.01mm$.

\section{Non-triviality of regular foliation}
\label{app:foliation}

Throughout this paper (eg. in subsection \ref{subsec:inconsistency}), we 
demand the regularity of time foliation as a physical condition. In this
appendix we shall see how significant this condition is. For this 
purpose, we consider a simple stellar solution in general relativity in
Painlev\'{e}-Gullstrand-like coordinate system and show that the time 
foliation is actually singular at the center. This makes it clear that
the regularity of time foliation is not a trivial but significant
condition. Based on this observation, we shall foresee what happens near
a stellar center in Ho\v{r}ava-Lifshitz gravity.

One of the simplest (and idealized) stellar solutions in general
relativity is that with uniform energy density. The explicit expression
can be found in textbooks, e.g.~\cite{wald}. By going to a
Painlev\'{e}-Gullstrand-like coordinate system, the solution is written
as 
\begin{eqnarray}
 ds^2 & = & 
  -dt^2+e^{-2\psi}(dr+\beta dt)^2+r^2d\Omega_2^2, \\
 \rho & = & \rho_0\ =\mbox{constant}, \\ 
 \frac{P}{\rho_0} & = & 
  \frac{(1-2M/R)^{1/2}-(1-2Mr^2/R^3)^{1/2}}
  {(1-2Mr^2/R^3)^{1/2}-3(1-2M/R)^{1/2}}, 
\end{eqnarray}
where
\begin{eqnarray}
 \psi &\!\! =\!\! & \left\{
  \begin{array}{lr}
   {1\over 8\pi} \log\left( -{1\over2}\left(1- 
      \sqrt{1-2MR \over 1-2Mr/R^3}    \right) \right)  &  (r<R) \\
   0 & (r>R)   \\
  \end{array}
 \right.\!\! , \\
\beta &\!\! =\!\! & \left\{
  \begin{array}{lr}
\sqrt{{8\pi\over 3}r^2 \rho_0 +e^{2\psi}-1} & (r<R) \\
    \sqrt{2M/r} & (r>R) \\
  \end{array}
 \right. ,\\
 M &\!\! =\!\! & {4\pi\over 3}R^3 \rho_0,
\end{eqnarray}
and $r=R$ is the surface of the star. For this solution, the Ricci
scalar of the three-dimensional spatial metric is 
\begin{equation}
 R=\frac{2}{r^2}(1 - e^{-2\psi} + 2 e^{-2\psi}r\frac{d\psi}{dr}),
\end{equation}
and diverges at the center. It is also easy to see that the extrinsic
curvature of constant-time hypersurfaces also diverges at the
center. Physical reason for these divergences is easy to understand. 
A metric with a constant lapse function is characterized by a congruence
of geodesics orthogonal to constant-time hypersurfaces. In general
relativity, a contracting congruence of geodesics forms caustics. This
is the physical reason why the three-dimensional spatial geometry and
the extrinsic curvature are singular at the center.

In general relativity, divergence of the extrinsic curvature and the
three-dimensional spatial curvature can be just a coordinate singularity.
Indeed, one can easily see that the four-dimensional geometry is regular
at the center of the above solution. Therefore, divergence at the
center is just a coordinate singularity and is not a problem for the
above solution in general relativity.

On the other hand, in Ho\v{r}ava-Lifshitz gravity, divergence of the
three-dimensional spatial curvature or/and the extrinsic curvature is a
physical singularity. Thus, the above solution with constant energy
density is not physically viable. Indeed, since the three-dimensional
spatial curvature diverges, higher spatial curvature terms become
important near the center and should change the behavior of the
solution. It is also expected that deviation of $\lambda$ from $1$
should also be important near the center. Those effects should generate 
``dark matter as integration constant'', which should develop would-be
caustics and bounce~\cite{DM}. What we expect is occurring near the
center is, thus, a sequence of microscopic would-be caustics and
bounces.


\begin{thebibliography}{1}
\bibitem{Horava:2009uw}
  P.~Horava,
  Phys.\ Rev.\  D {\bf 79}, 084008 (2009)
  [arXiv:0901.3775 [hep-th]].

\bibitem{Horava:2009if}
  P.~Horava,
  Phys.\ Rev.\ Lett.\  {\bf 102}, 161301 (2009)
  [arXiv:0902.3657 [hep-th]].

\bibitem{cos}

%
  T.~Takahashi and J.~Soda,
  Phys.\ Rev.\ Lett.\  {\bf 102}, 231301 (2009)
  [arXiv:0904.0554 [hep-th]].
%
  G.~Calcagni,
  arXiv:0904.0829 [hep-th].
%
  E.~Kiritsis and G.~Kofinas,
  arXiv:0904.1334 [hep-th].
%
  S.~Mukohyama,
  JCAP {\bf 0906}, 001 (2009)
  [arXiv:0904.2190 [hep-th]].
%
  Y.~S.~Piao,
  arXiv:0904.4117 [hep-th].
%
  X.~Gao,
  arXiv:0904.4187 [hep-th].
%
  S.~Mukohyama, K.~Nakayama, F.~Takahashi and S.~Yokoyama,
  Phys.\ Lett.\  B {\bf 679}, 6 (2009)
  [arXiv:0905.0055 [hep-th]].
%
  S.~Kalyana Rama,
  Phys.\ Rev.\  D {\bf 79}, 124031 (2009)
  [arXiv:0905.0700 [hep-th]].
%
  B.~Chen, S.~Pi and J.~Z.~Tang,
  arXiv:0905.2300 [hep-th].
%
  E.~N.~Saridakis,
  arXiv:0905.3532 [hep-th].
%
  X.~Gao, Y.~Wang, R.~Brandenberger and A.~Riotto,
  arXiv:0905.3821 [hep-th].
%
  M.~Minamitsuji,
  arXiv:0905.3892 [astro-ph.CO].
%
  A.~Wang and Y.~Wu,
  JCAP {\bf 0907}, 012 (2009)
  [arXiv:0905.4117 [hep-th]].
%
  S.~Nojiri and S.~D.~Odintsov,
  arXiv:0905.4213 [hep-th].
%
  Y.~F.~Cai and E.~N.~Saridakis,
  arXiv:0906.1789 [hep-th].
%
  Y.~F.~Cai and X.~Zhang,
  Phys.\ Rev.\  D {\bf 80}, 043520 (2009)
  [arXiv:0906.3341 [astro-ph.CO]].
%
  M.~i.~Park,
  arXiv:0906.4275 [hep-th].
%
  K.~Yamamoto, T.~Kobayashi and G.~Nakamura,
  arXiv:0907.1549 [astro-ph.CO].
%
  A.~Wang and R.~Maartens,
  arXiv:0907.1748 [hep-th].
%
  Y.~Lu and Y.~S.~Piao,
  arXiv:0907.3982 [hep-th].
%
  T.~Kobayashi, Y.~Urakawa and M.~Yamaguchi,
  arXiv:0908.1005 [astro-ph.CO].
%
  S.~Koh,
  arXiv:0907.0850 [hep-th].
%
  C.~Bogdanos and E.~N.~Saridakis,
  arXiv:0907.1636 [hep-th].
%
  C.~Appignani, R.~Casadio and S.~Shankaranarayanan,
  arXiv:0907.3121 [hep-th].
%
  M.~R.~Setare,
  arXiv:0909.0456 [hep-th].
%
  S.~Maeda, S.~Mukohyama and T.~Shiromizu,
  arXiv:0909.2149 [astro-ph.CO].
%
  S.~Carloni, E.~Elizalde and P.~J.~Silva,
  arXiv:0909.2219 [hep-th].
%
C.~Ding, S.~Chen and J.~Jing,
  arXiv:0909.2490 [gr-qc].
%
  P.~Wu and H.~W.~Yu,
  arXiv:0909.2821 [gr-qc].
%
  G.~Leon and E.~N.~Saridakis,
  arXiv:0909.3571 [hep-th].
%
  C.~G.~Boehmer and F.~S.~N.~Lobo,
  arXiv:0909.3986 [gr-qc].
%
  A.~Wang, D.~Wands and R.~Maartens,
  arXiv:0909.5167 [hep-th].
%
  B.~Chen, S.~Pi and J.~Z.~Tang,
  arXiv:0910.0338 [hep-th].
%
  M.~i.~Park,
  arXiv:0910.1917 [hep-th].
%
  S.~Dutta and E.~N.~Saridakis,
  arXiv:0911.1435 [hep-th].


\bibitem{DM}
  S.~Mukohyama,
  Phys.\ Rev.\  D {\bf 80}, 064005 (2009)
  [arXiv:0905.3563 [hep-th]].
%
  S.~Mukohyama,
  arXiv:0906.5069 [hep-th].
%

\bibitem{BH}

  H.~Lu, J.~Mei and C.~N.~Pope,
  Phys.\ Rev.\ Lett.\  {\bf 103}, 091301 (2009)
  [arXiv:0904.1595 [hep-th]].
%
  E.~O.~Colgain and H.~Yavartanoo,
  JHEP {\bf 0908}, 021 (2009)
  [arXiv:0904.4357 [hep-th]].
%
  R.~G.~Cai, L.~M.~Cao and N.~Ohta,
  Phys.\ Rev.\  D {\bf 80}, 024003 (2009)
  [arXiv:0904.3670 [hep-th]].
%
  R.~G.~Cai, Y.~Liu and Y.~W.~Sun,
  JHEP {\bf 0906}, 010 (2009)
  [arXiv:0904.4104 [hep-th]].
%
  A.~Ghodsi,
  arXiv:0905.0836 [hep-th].
%
  Y.~S.~Myung and Y.~W.~Kim,
  arXiv:0905.0179 [hep-th].
%
  A.~Kehagias and K.~Sfetsos,
  Phys.\ Lett.\  B {\bf 678}, 123 (2009)
  [arXiv:0905.0477 [hep-th]].
%
  R.~G.~Cai, L.~M.~Cao and N.~Ohta,
  arXiv:0905.0751 [hep-th].
%
  Y.~S.~Myung,
  Phys.\ Lett.\  B {\bf 678}, 127 (2009)
  [arXiv:0905.0957 [hep-th]].
%
  A.~Ghodsi and E.~Hatefi,
  arXiv:0906.1237 [hep-th].
%
  R.~A.~Konoplya,
  Phys.\ Lett.\  B {\bf 679}, 499 (2009)
  [arXiv:0905.1523 [hep-th]].
%
  H.~W.~Lee, Y.~W.~Kim and Y.~S.~Myung,
  arXiv:0907.3568 [hep-th].
%
  Y.~S.~Myung,
  arXiv:0908.4132 [hep-th].
%
  E.~Kant, F.~R.~Klinkhamer and M.~Schreck,
  arXiv:0909.0160 [hep-th].
%
  J.~Z.~Tang and B.~Chen,
  arXiv:0909.4127 [hep-th].
%
N.~Varghese and V.~C.~Kuriakose,
  arXiv:0909.4944 [gr-qc].
%
  M.~Jamil, E.~N.~Saridakis and M.~R.~Setare,
  arXiv:0910.0822 [hep-th].
%
  R.~G.~Cai and N.~Ohta,
  arXiv:0910.2307 [hep-th].
  %
  D.~Y.~Chen, H.~Yang and X.~T.~Zu,
  arXiv:0910.4821 [gr-qc].
%
  E.~Kiritsis and G.~Kofinas,
  arXiv:0910.5487 [hep-th].
%
  D.~Capasso and A.~P.~Polychronakos,
  arXiv:0911.1535 [hep-th].
%
  T.~Harada, U.~Miyamoto and N.~Tsukamoto,
  arXiv:0911.1187 [gr-qc].


\bibitem{sol}

  T.~Harko, Z.~Kovacs and F.~S.~N.~Lobo,
  Phys.\ Rev.\  D {\bf 80}, 044021 (2009)
  [arXiv:0907.1449 [gr-qc]].
%
  T.~Harko, Z.~Kovacs and F.~S.~N.~Lobo,
  arXiv:0908.2874 [gr-qc].
%
  C.~Ding, S.~Chen and J.~Jing,
  arXiv:0909.2490 [gr-qc].
%
  L.~Iorio and M.~L.~Ruggiero,
  arXiv:0909.2562 [gr-qc].
%
  J.~Z.~N.~Tang and B.~Chen,
  arXiv:0909.4127 [hep-th].
%
  L.~Iorio and M.~L.~Ruggiero,
  arXiv:0909.5355 [gr-qc].


\bibitem{others}
%
  T.~P.~Sotiriou, M.~Visser and S.~Weinfurtner,
  Phys.\ Rev.\ Lett.\  {\bf 102}, 251601 (2009)
  [arXiv:0904.4464 [hep-th]].
%
  T.~P.~Sotiriou, M.~Visser and S.~Weinfurtner,
  JHEP {\bf 0910}, 033 (2009)
  [arXiv:0905.2798 [hep-th]].
%
  T.~Nishioka,
  arXiv:0905.0473 [hep-th].
%
  Y.~S.~Myung,
  Phys.\ Lett.\  B {\bf 679}, 491 (2009)
  [arXiv:0907.5256 [hep-th]].
%
  Y.~S.~Myung,
  Phys.\ Lett.\  B {\bf 681}, 81 (2009)
  [arXiv:0909.2075 [hep-th]].
%
  I.~Cho and G.~Kang,
  arXiv:0909.3065 [hep-th].
%
  D.~Blas, O.~Pujolas and S.~Sibiryakov,
  arXiv:0909.3525 [hep-th].
%
  T.~Suyama,
  arXiv:0909.4833 [hep-th].
%
  D.~Capasso and A.~P.~Polychronakos,
  arXiv:0909.5405 [hep-th].
%
  A.~Papazoglou and T.~P.~Sotiriou,
  arXiv:0911.1299 [hep-th].



\bibitem{Charmousis:2009tc}
  C.~Charmousis, G.~Niz, A.~Padilla and P.~M.~Saffin,
  JHEP {\bf 0908}, 070 (2009)
  [arXiv:0905.2579 [hep-th]].

\bibitem{Li:2009bg}

  M.~Li and Y.~Pang,
  arXiv:0905.2751 [hep-th].
%


\bibitem{Fierz:1939ix}
  M.~Fierz and W.~Pauli,
  Proc.\ Roy.\ Soc.\ Lond.\  A {\bf 173}, 211 (1939).

\bibitem{Vainshtein:1972sx}
  A.~I.~Vainshtein,
  Phys.\ Lett.\  B {\bf 39}, 393 (1972).

\bibitem{wald}
 R.~M.~Wald,
  "General Relativity," 
  The University of Chicago Press (1984).

\bibitem{Wang:2009yz}
  A.~Wang and R.~Maartens,
  arXiv:0907.1748 [hep-th].
%
  D.~Blas, O.~Pujolas and S.~Sibiryakov,
  arXiv:0909.3525 [hep-th].
%
  K.~Koyama and F.~Arroja,
  arXiv:0910.1998 [hep-th].


\end{thebibliography}
\end{document}